\documentclass[12pt]{article}

\usepackage{graphics,graphicx,fullpage,natbib,multirow}
\usepackage{amsmath,amssymb,verbatim,epsfig}
\usepackage[dvipsnames,usenames]{color}

\newtheorem{defin}{\bf Definition}

\def\E{\mbox{E}}

\def\d{\mbox{d}}

\def\bn{{\bf n}}

\def\bx{{\bf x}}
\def\by{{\bf y}}

\newcommand{\RB}{\mathbb{R}}

\begin{document}

\baselineskip=24pt

\title{\bf Leaf clustering using circular densities}
\author{{\sc Luis E. Nieto-Barajas} \\[2mm]
{\sl Department of Statistics, ITAM, Mexico} \\[2mm]
{\small luis.nieto@itam.mx}}
\date{}
\maketitle

\begin{abstract}
In the biology field of botany, leaf shape recognition is an important task. One way of characterising the leaf shape is through the centroid contour distances (CCD). Each CCD path might have different resolution, so normalisation is done by associating each contour to a circular density. Densities are rotated by subtracting the mean or mode preferred direction. Distance measures between densities are used to produce a hierarchical clustering method to cluster the leaves. We illustrate our approach with a motivating small dataset as well as a larger dataset. 
\end{abstract}

\vspace{0.2in} \noindent {\sl Keywords}: Circular density, center contour distance, hierarchical clustering, time series clustering.

\section{Introduction}
\label{sec:intro}

The branch of biology that studies plants is botany, also called phytology. Historically, plant kingdom contained all living things that were not animals, however, nowadays fungi and algae are excluded \citep{mauseth:19}. There are more that three hundred thousand species of plants, so species identification has been an important task to define such a taxonomy of plants. 

There are several ways of characterising a plant. For plants with leaves, most of the proposals concentrate on the features of their leaves, such as shape, vein, colour and texture. One way of characterising the shape is through a centroid contour distance (CCD) \citep{cope&al:12}. CCD is a contour-based approach that consists in segmenting the silhouette of the leaf in a two dimensional image, and for each pixel, compute its distance to the centroid. CCD depends on the resolution of the image, so different leaves might have different number of CCD points recorded. 

CCD data has to be normalised in two aspects. First, the same leaf might lead to  different CCD representations based on the scale (closeness of the camera to the leaf), second, the starting point recorded for each leaf might not be the same, so reported data is subject to a rotation. Authors have proposed different ways of achieving this normalisation. 

\cite{wang&al:00} for example, do a subsampling of the contour points to get rid of the scale and do a shifting of the curve wrappedly, but to minimize the number of rotations they also perform a thinning to find a skeleton and simplify the computations in the matching process. \cite{teng&al:09} simply divide by the maximum value to get rid of the scale and use a circular shift together with a similarity measure to match rotations. On the other hand, \cite{kadir&al:11} uses normalised polar Fourier coefficients to get rid of the scale and rotation and they further use other shape, colour, vein and texture features in a probabilistic neural network to classify. Finally, \cite{hasim&al:16} did a manual rotation to ensure that all leaves are in a vertical position and do not propose any scale correction arguing that their classification method is invariant under scale changes. 

If all leaves have the same number of CCD points, are not rotated and share the same resolution, data is treated as time series and clustering of time series has been studied, for example by \cite{nieto&contreras:14}. However, this is not the typical scenario for CCD data. 

On the other hand, when each leaf is characterised by other high dimensions, \cite{ding&al:16} performed leaf clustering reducing the dimensionality via an sparse subspace clustering. 

In this article we propose to characterise the CCD data as circular densities. By doing so the normalisation process is straightforward. We get rid of the scale by normalising the density and get rid of the rotation by substracting the mean or mode of the density.  Distances between densities are then used to define a dissimilarity matrix and perform a hierarchical clustering to cluster the leaves. The task of clustering densities has also been studied by \cite{phamtoan&vovan:20}. 

The contents of the rest of the paper is as follows: In Section \ref{sec:ccd} we characterise the CCD data as circular densities and propose an ingenious normalisation process to get rid of scale and rotation problems. In Section \ref{sec:clustering} we propose four types of distances to perform a hierarchical clustering. Data analysis of our motivating dataset is carried out along Sections \ref{sec:ccd} and \ref{sec:clustering}. We later perform an analysis of a larger dataset in Section \ref{sec:osu}. We finally conclude in Section \ref{sec:concl}.

\section{CCD data as circular densities}
\label{sec:ccd}

Let us assume that each leaf $i$ is characterised by a sequence of size $n_i$ of center contour distances (CCD) $\by_i=(y_{i1},\ldots,y_{i,n_i})$, where $y_{ij}$ is the CCD at location $x_{ij}$ for $j=1,\ldots,n_i$ and $i=1,\ldots,m$ with $m$ the number of leaves in the sample. Usually, locations $x_{ij}$ are not available but it is assumed that the $n_i$ measurements $y_{ij}$ cover the entire contour of the leaf. We will assume that locations are uniformly spread angles around the circle $(0,2\pi]$ in radians, therefore $x_{i,j}=2\pi j/n_i$.

We use $\by_i$ and $\bx_i=(x_{i1},\ldots,x_{i,n_i})$ to characterise each leaf by a circular density $f_i(t)$ \citep[e.g.][]{jamma&gupta:01} in the following way:
\begin{equation}
\label{eq:ft1}
f_i(t)=c_i\sum_{j=1}^{n_i}y_{ij}I_{(x_{i,j-1},x_{ij}]}(t),
\end{equation}
for $t\in(0,2\pi]$ where $x_{i0}=0$ and $c_i$ is a normalising constant such that $\int f_i(t)\d t=1$. It is straightforward to show that $c_i=1/(2\pi\bar{y_i})$ with $\bar{y_i}=(1/n_i)\sum_{j=1}^{n_i}y_{ij}$. By defining each leaf as a density, as in \eqref{eq:ft1}, we get rid of the scale of the CCD data that is not relevant for the clustering. 

Another confounding aspect of the CCD data is the rotation, the first available measure $y_{i1}$ might not be made at the same location of the leaf for different $i$'s. Since we have arbitrarily defined the locations $x_{ij}$ around the circle, we suggest to rotate the densities in two ways. If we let $T_i$ be a random variable with probability density $f_i(t)$, then two location parameters are the mean and the mode. The mean preferred angle $\mu_i$, for circular densities, is defined via the first trigonometric moments as $\mu_i=\arctan^*(\beta_i/\alpha_i)$ with $\alpha_i=\E(\cos T_i)$ and $\beta_i=\E(\sin T_i)$. Here, $\arctan^*(\beta_i/\alpha_i)$ is an slightly modified function defined as $\arctan(\beta_i/\alpha_i)$ if $\beta_i>0$ and $\alpha_i>0$, $\arctan(\beta_i/\alpha_i)+\pi$ if $\alpha_i<0$, and $\arctan(\beta_i/\alpha_i)+2\pi$ if $\beta_i<0$ and $\alpha_i>0$. On the other hand, the mode angle $\nu_i$ is defined, as usual, as the value of the support where the density is maximum. 

Therefore, we define the rotated random variable $T_i^*=T_i-\lambda_i$, which has density
\begin{equation}
\label{eq:ft2}
f_i^*(t)=\frac{1}{2\pi\bar{y_i}}\sum_{j=1}^{n_i}y_{ij}I_{(x_{i,j-1}^*,x_{ij}^*]}(t),
\end{equation}
where $x_{ij}^*=x_{ij}-\lambda_i$ and $\lambda_i$ is either the mean $\mu_i$ or the mode $\nu_i$. 

In principle, circular densities are defined for $t\in\RB$ and must satisfy the property $f_i^*(t+2\pi)=f_i^*(t)$. Density \eqref{eq:ft2} is a perfectly defined circular density, which is better depicted around a circle. However, we are used to plot densities in a Cartesian system. Doing so might not be direct since rotated densities might have negative $x_{ij}^*$ values. Considering the circular property, we can force the support to be constrained to $t\in(0,2\pi]$ by replacing negative $x_{ij}^*$ values by $x_{ij}^*+2\pi$. 

To illustrate how our normalisation and rotation proposal works, let us consider a dataset of size $m=10$, that consists of marine plants for which the type is already known. Three of them are FRS, two MOS, three AVS and two AGS. Unfortunately the meaning of these names is not known for us. Knowing the class in advance is not the typical setting for clustering, but will help us understand how our procedure works. 

Available information consists of CCD measurements $\by_i$ of unequal sizes $n_i$, for $i=1,\ldots,m$. In particular, vector sizes are $$\bn=(3603, 3392, 3293, 2649, 2270, 3633, 3623, 3791, 3864, 3035).$$ 

\begin{figure}
\centerline{\includegraphics[scale=0.47]{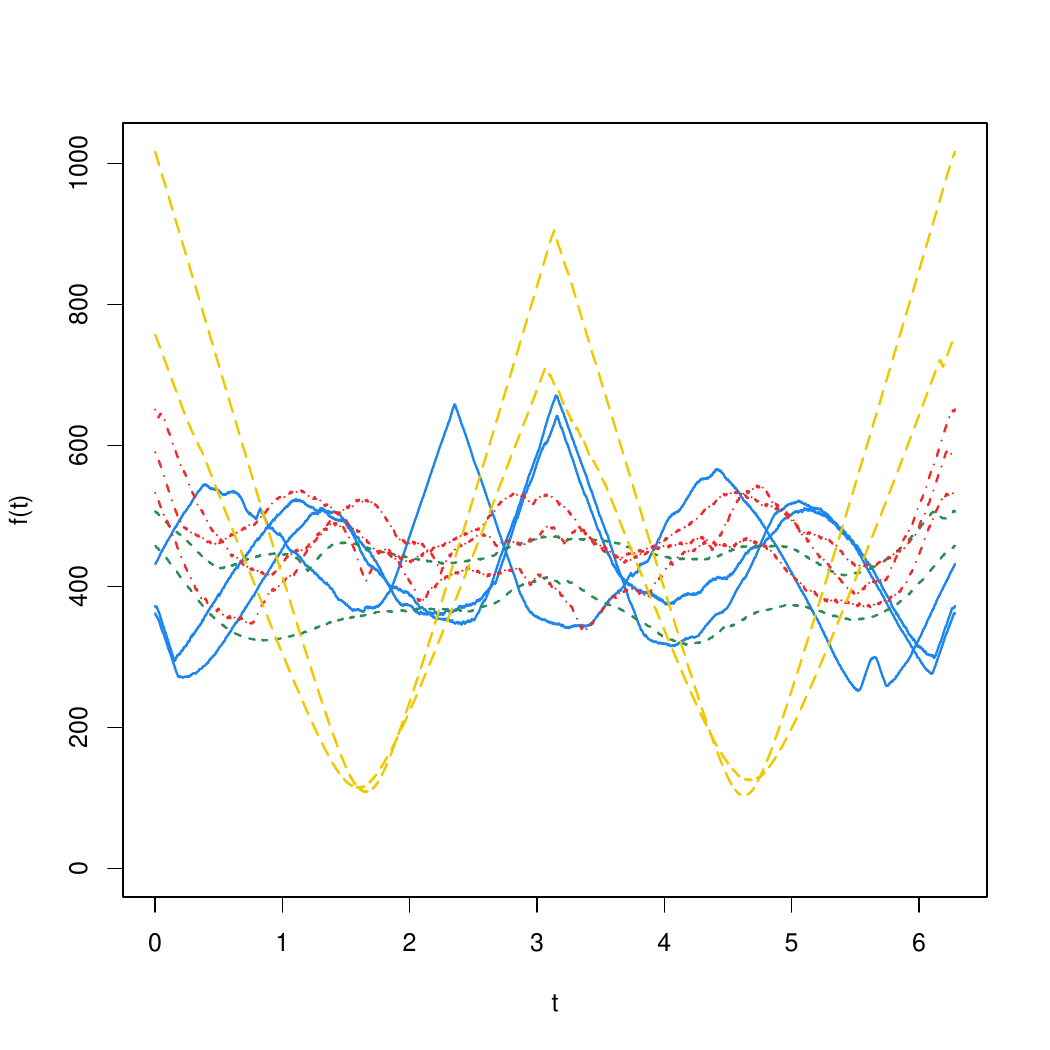}
\includegraphics[scale=0.47]{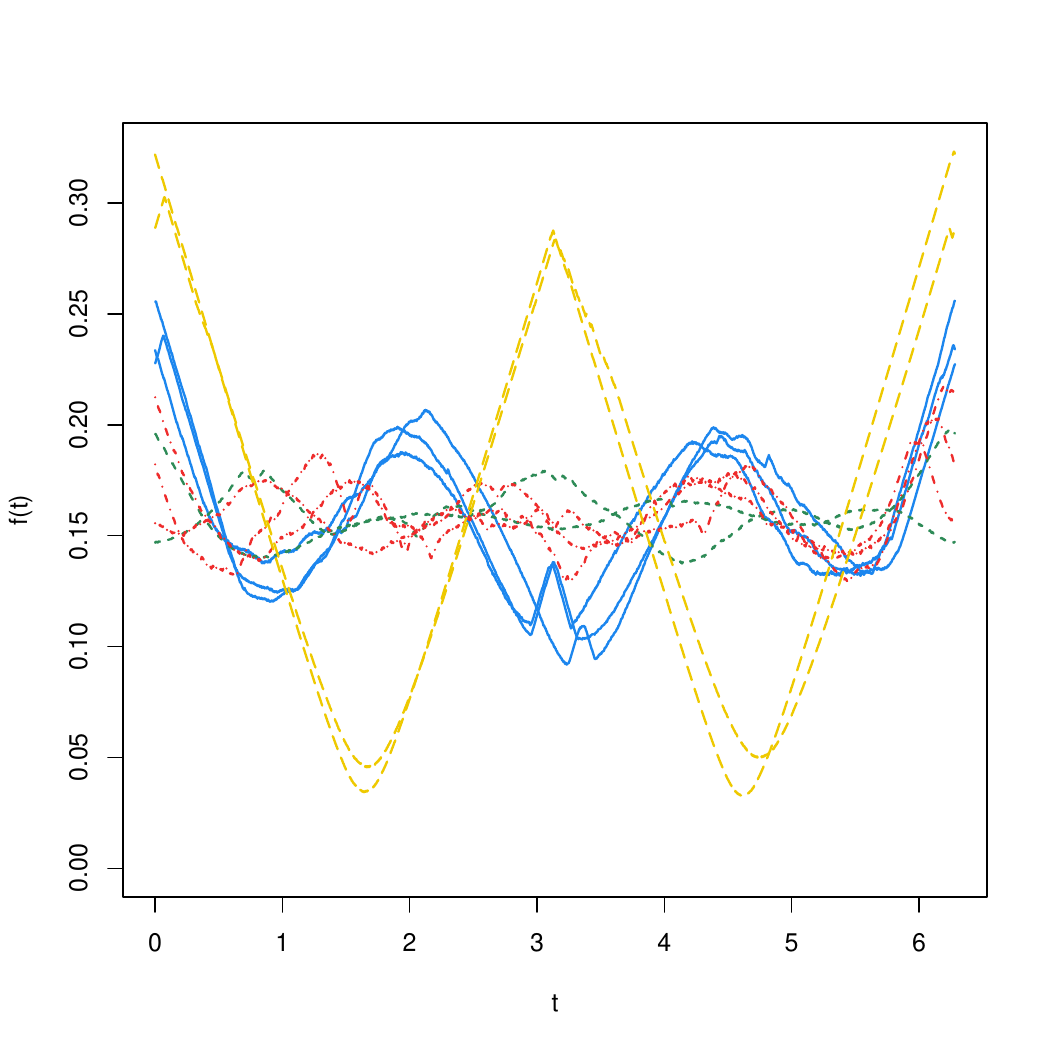}}
\vspace{-2mm}
\caption{{\small Ten leaves CCD data as circular densities. Without normalisation (left) and with normalisation and mean rotation (right).}}
\label{fig:dens}
\end{figure}

In Figure \ref{fig:dens} we show the unnormalised densities \eqref{eq:ft1}, i.e. assuming $c_i=1$ in the left panel, and normalised and mean rotated densities \eqref{eq:ft2}, in the right panel. We have used different colour and line type for the four plants types. First note that in the left panel, vertical scale goes from $0$ to $1000$ and lines have different amplitudes, AVS leaves (red dotted-dashed lines) have small amplitude, whereas AGS leaves (yellow dashed lines) have huge amplitudes. 
For the FRS plants (blue solid lines) the raw data did not show a scale problem but a rotation in one of the three leaves, however for the AGS plants (yellow dashed lines), rotation is not the problem but a mismatch in the scales between the two leaves. Both problems were resolved with our proposal in the right panel. 

Apart from representing each leaf as a circular density, we can plot the actual leaves by considering that the pairs $(x_{ij},c_iy_{i,j})$ in \eqref{eq:ft1} or $(x_{ij}^*,c_iy_{i,j})$ in \eqref{eq:ft2}, for $j=1,\ldots,n_i$ and $i=1,\ldots,n$, are points in polar coordinates where the abscissa corresponds to the angle and the ordinate to the radius or resultant length. We can then transform the polar coordinates to Cartesian coordinates by doing $u_{ij}=c_iy_{ij}\cos(x_{ij})$ and $v_{ij}=c_iy_{ij}\sin(x_{ij})$ and plot the pairs $(u_{ij},v_{ij})$ for $j=1,\ldots,n_i$ and $i=1,\ldots,n$. 

\begin{figure}
\centerline{\includegraphics[scale=0.7]{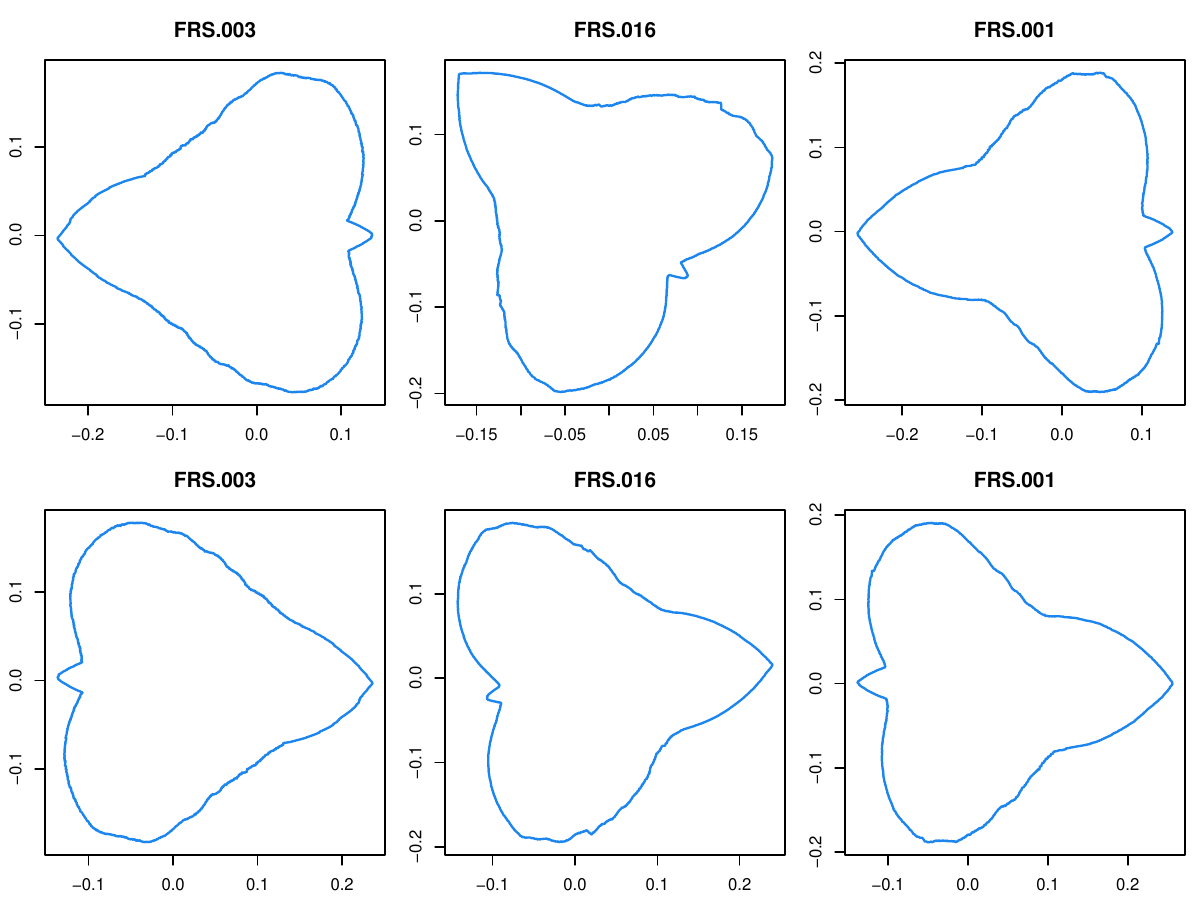}}
\caption{{\small FRS leaves plot. Before rotation (top) and after mean rotation (bottom). The dot represents the center of the leaf.}}
\label{fig:leaves01}
\end{figure}

\begin{figure}
\centerline{\includegraphics[scale=0.7]{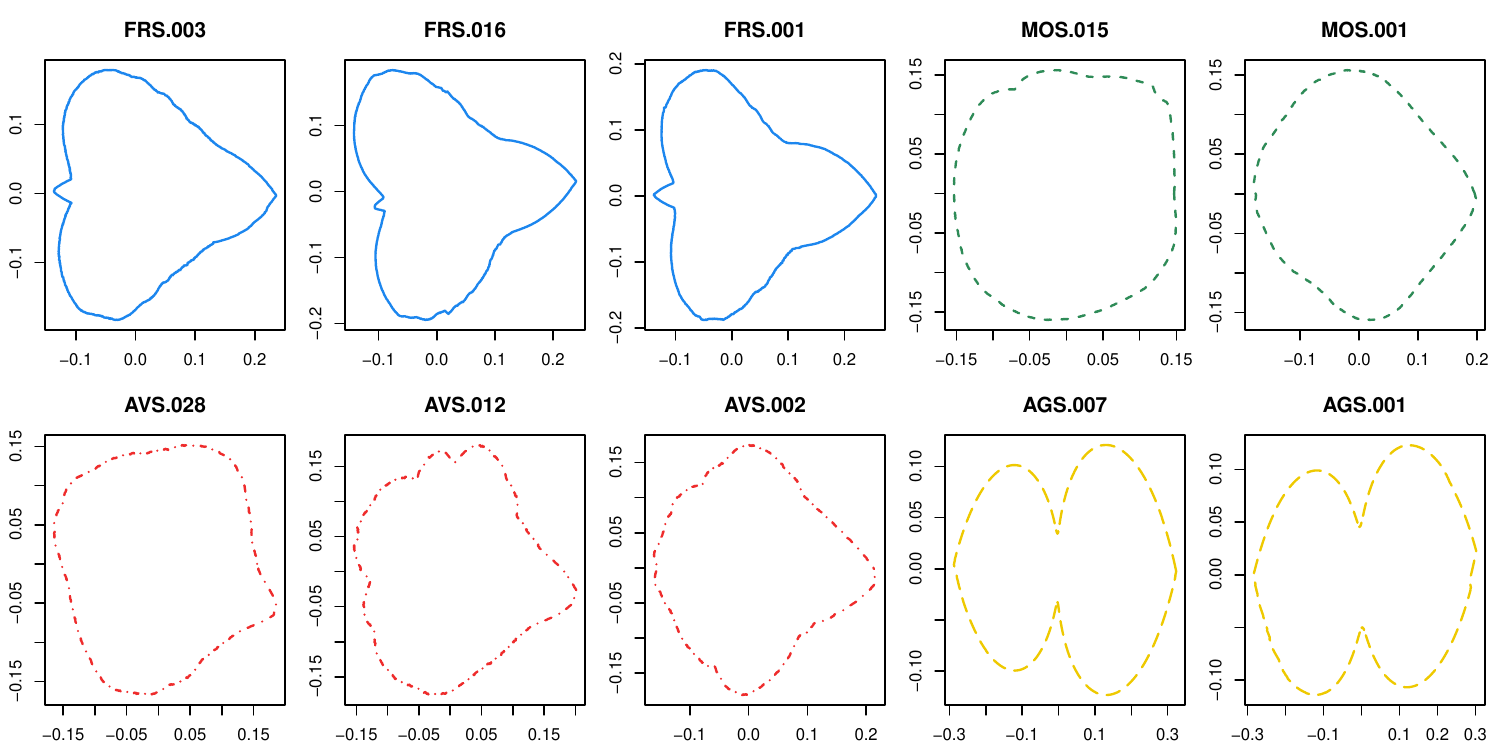}}
\caption{{\small Ten leaves plot after mean rotation.}}
\label{fig:leaves1a}
\end{figure}

In Figure \ref{fig:leaves01} we show the three leaves of FRS type. In the top line we show the original orientation \eqref{eq:ft1}, where leaves FRS.003 and FRS.001 (first and third) have a horizontal orientation, whereas leaf FRS.016 appears in diagonal. After implementing our mean rotation correction \eqref{eq:ft2}, shown in the second row, all three leaves are facing towards the west. The center of each leaf, from where the CCD is measur, is the origin represented with a dot in the plot. 
Finally, Figure \ref{fig:leaves1a} presents all ten leaves after normalisation, scale and mean rotation corrections.

\section{Hierarchical clustering}
\label{sec:clustering}

So far, each leaf $i$, originally defined by its CCD measurements $\by_i$, has been characterised by the circular density $f_i^*(t)$ given in \eqref{eq:ft2}. To perform a hierarchical clustering \citep[e.g.][]{johnson&wichern:07} we need a similarity/dissimilarity measure between any two densities $f_i^*(t)$ and $f_k^*(t)$, say $D(f_i^*,f_k^*)$. There are several options to measure distances between densities, here we consider three of them that include the $L_1$, the total variation and the Hellinger distances: 
\begin{align}
\label{eq:dist}
\nonumber
D_1(f_i^*,f_k^*)&=\int_0^{2\pi}|f_i^*(t)-f_k^*(t)| \d t \\
\nonumber
D_2(f_i^*,f_k^*)&=\sup_{t\in(0,{2\pi}]}|f_i^*(t)-f_k^*(t)| \\
\nonumber
D_3(f_i^*,f_k^*)&=\int_0^{2\pi}\left[\{f_i^*(t)\}^{1/2}-\{f_k^*(t)\}^{1/2}\right]^2 \d t
\end{align}

Alternatively, each circular density $f_i^*(t)$ can be entirely characterised by its trigonometric moments, these are $\alpha_{i}(p)=\E\{\cos(pT_i)\}$ and $\beta_{i}(p)=\E\{\sin(pT_i)\}$ for $p=1,2,\ldots$. If we only consider the first $2r$ moments, we can use euclidean distance as dissimilarity measure. We call this $D_4(f_i^*(t),f_k^*(t))$. This latter approach is similar to \cite{kadir&al:11} in the sense that they use normalised polar Fourier coefficients, but we do not require further normalisation as they do. 

Once we have the distance function, we have to decide the link method to define the distances between clusters. In our experience, complete linkage provides good clusterings since it minimises the maximum distance within clusters \citep{glasbey:87}. We will use this link. 

\begin{figure}
\centerline{\includegraphics[scale=0.47]{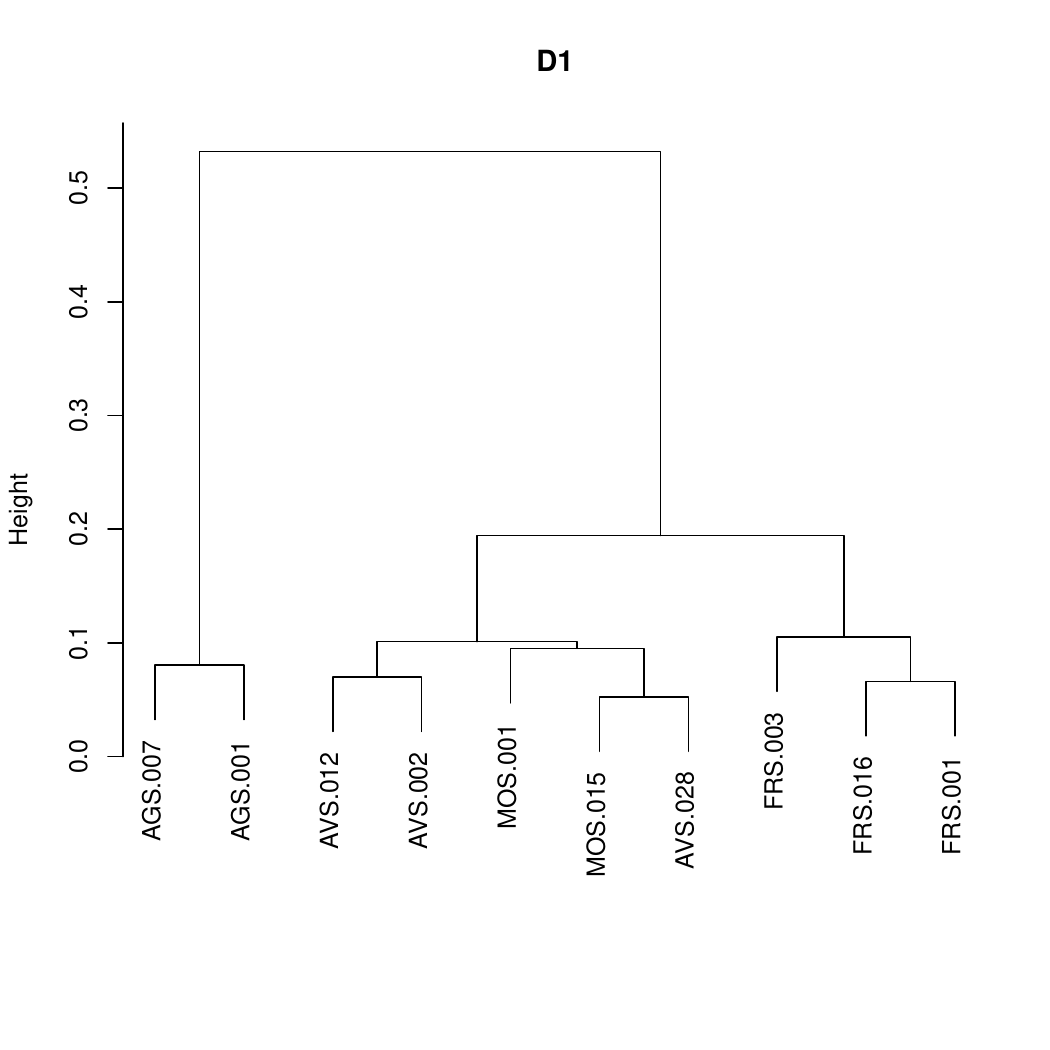}
\includegraphics[scale=0.47]{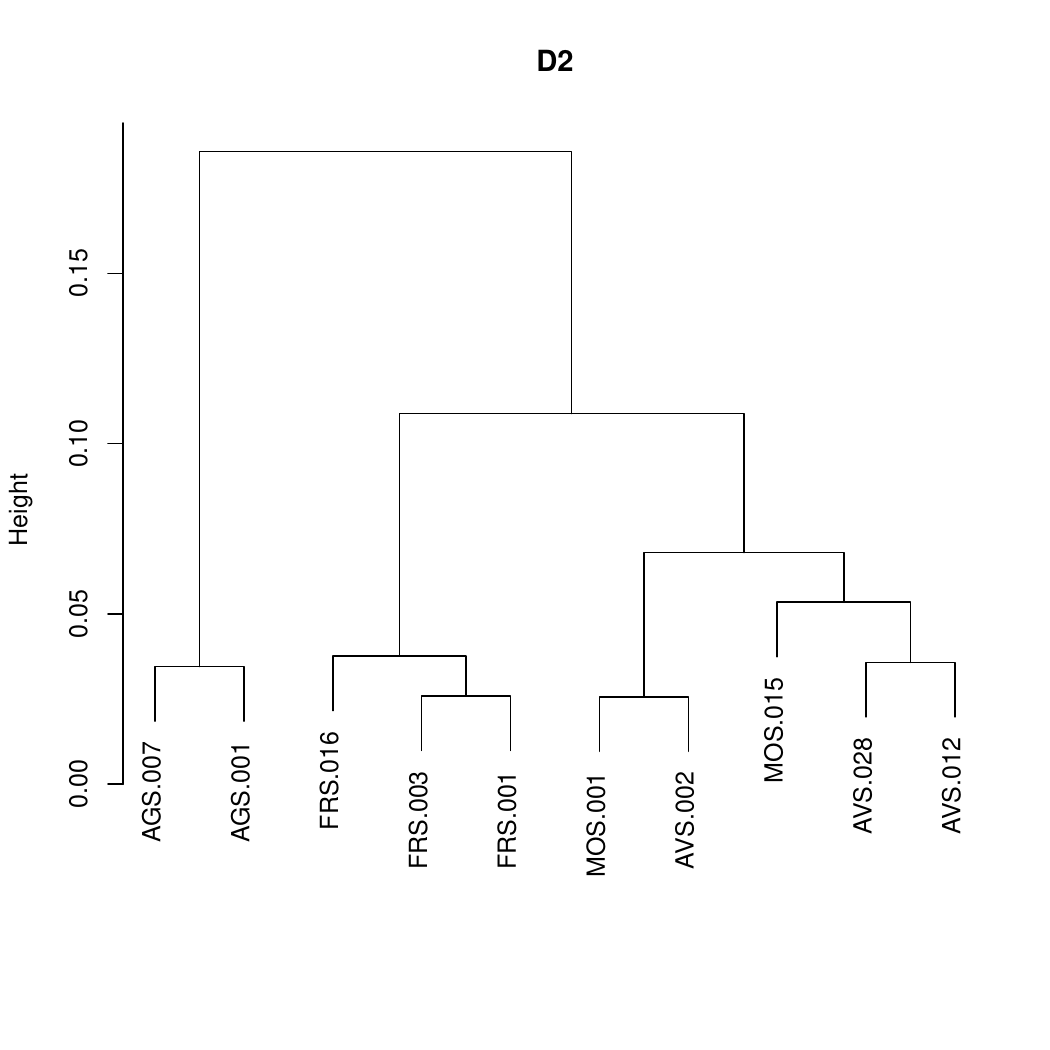}}
\centerline{\includegraphics[scale=0.47]{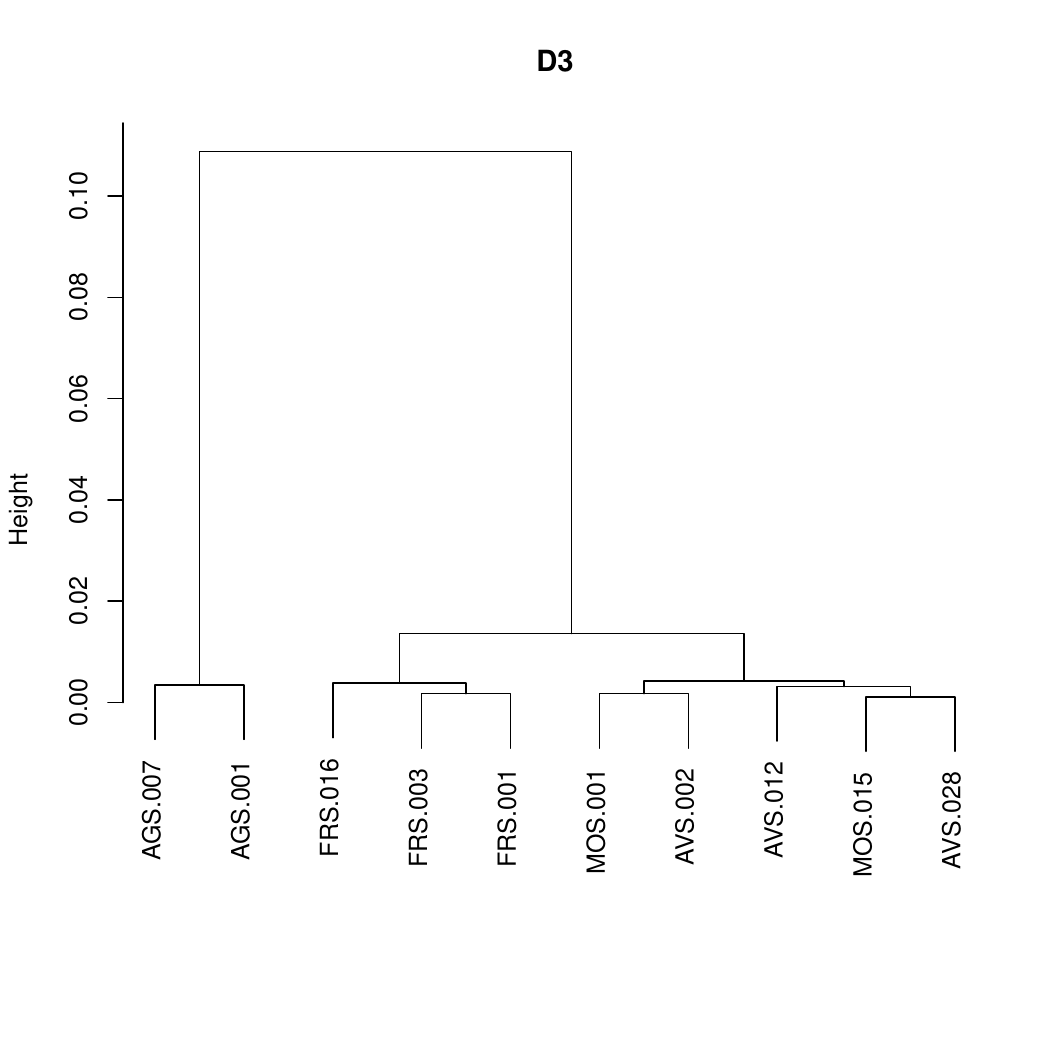}
\includegraphics[scale=0.47]{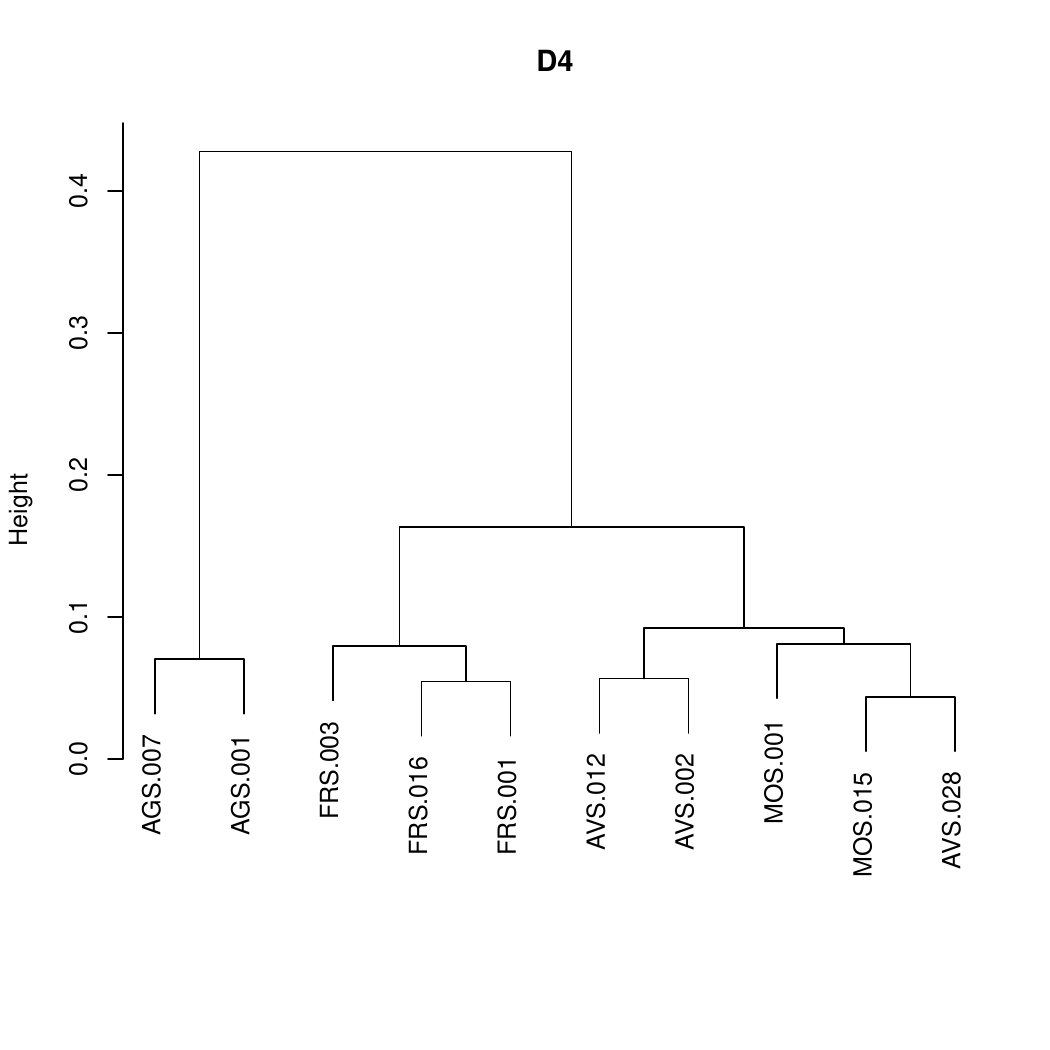}}
\vspace{-2mm}
\caption{{\small Ten leaves dataset. Hierarchical clustering dendrograms for the four distances.}}
\label{fig:dendro}
\end{figure}

Figure \ref{fig:dendro} shows the dendrograms obtained for the ten leaves dataset and for the four distances. In particular, for distance $D_4$ we used $r=5$ to characterise the densities. We note that the vertical axes show different scales due to the different distances used. Recall that a longer vertical line in a dendrogram means that the individuals below are more dissimilar to the others. 

In all cases, the group of the two AGS plants is well identified. The group of three FRS plants is also identified by the four dendrograms, however, it is better separated (longer line) by the total variation distance ($D_2$), followed by the $L_1$ ($D_1$) and trigonometric moments ($D_4$) distances. The least clear job is made by the Hellinger ($D_3$) distance. Regarding plants MOS and AVS, all four distances mixed up the identification of the types. This is understandable since green and red densities in Figure \ref{fig:dens} (right panel) are overlapped, and green and red leaves pictures in Figure \ref{fig:leaves1a} look very similar. $L_1$ ($D_1$) distance joins the two MOS leaves with leaf AVS.028 and put together the other two AVS.012 and AVS.002 leaves. Total variation ($D_2$) distance puts together MOS.015, AVS.028 and AVS.012 in one group and MOS.001 and AVS.002 in another. Hellinger ($D_3$) distance forms the same groups as the total variation distance, but perhaps in a less clear way (shorter vertical lines). Finally, trigonometric moments combined with euclidean distances ($D_4$) produces the same groups as the $L_1$ distance. 

In summary, $D_1$ \& $D_4$ and $D_2$ \& $D_3$ produce the sample clusterings. In terms of errors, $D_1$ \& $D_4$ only misplace one leaf, whereas $D_2$ \& $D_3$ misplace two leaves. Visually judging the dendrograms, we believe that $D_1$ \& $D_4$ are equally good and $D_2$ is preferred to $D_3$ in terms of clarity to identify the groups.

\section{OSU Leaf dataset}
\label{sec:osu}

In this section we present an analysis of a more comprehensive dataset. The time series machine learning website contains data for time series clustering and classification. The Oregon State University (OSU) leaf dataset, available at \texttt{
https://www.timeseriesclassifica tion.com/description.php?Dataset=OSULeaf}
consists of one dimensional outlines (CCDs) of leaves. The dataset contains $m=200$ leaves characterised by $n_i=428$ datapoints for $i=1,\ldots,m$. Before analysing these data, we removed the first observation of each series because it was associated to the stem, and added the absolute value of the minimun of the remaining observations to produce positive values. 

The website mentions that the series were obtained by color image segmentation and boundary extraction, in the anti-clockwise direction, from digitized leaf images of six classes: Acer Circinatum, Acer Glabrum, Acer Macrophyllum, Acer Negundo, Quercus Garryana and Quercus Kelloggii. However no classification identification was provided for any of the $m=200$ leaves. Therefore we perform a clustering, using our proposal, and define $L=6$ groups with the hope that they coincide with the original classes. 

We first defined the densities as in \eqref{eq:ft2}, one for each of the the $200$ leaves. Since the resulting densities show high multimodality, we rotated them with both, the mean and the mode. For each of the two types of rotation, we computed the four distances mentioned in Section \ref{sec:clustering} with $r=3$ for distance $D_4$. To compare among the different clusterings, we define a heterogeneity measure (HM), similar in spirit to the one defined in \cite{nieto&contreras:14}, given by
$$HM(G_1,\ldots,G_L)=\sum_{l=1}^L\frac{1}{n_l-1}\sum_{i<k\in G_l}\int_0^{2\pi}\left\{f_i(t)-f_k(t)\right\}^2\d t,$$
where $G_l$ denotes group $l$, for $l=1,\ldots,L$. 

The heterogeneity measures as well as the number of leaves in each of the six groups, obtained by the different distances, are given in Table \ref{tab:hm}. We first note that the group sizes vary a lot depending on the rotation and distance used. Comparing the $HM$'s, smaller values are obtained when the rotation is performed with the mode, as compared to the mean. The least heterogeneous clustering is obtained with $L_1$ distance $D_1$ and with mode rotation. The second best clustering is obtained with the Hellinger distance $D_3$ and with mode rotation. 

\begin{figure}
\centerline{\includegraphics[scale=0.9]{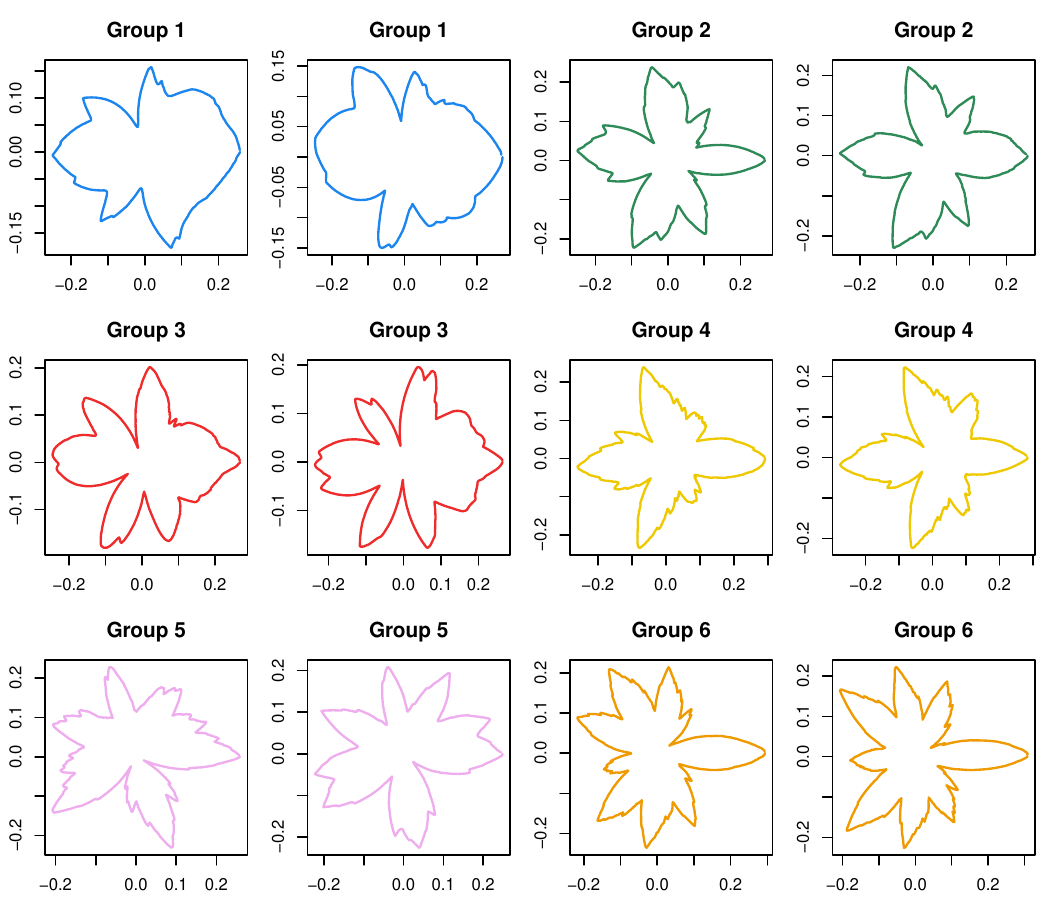}}
\caption{{\small OSU dataset. Plot of two leaves for each of the six clusters obtained.}}
\label{fig:osusum}
\end{figure}

To visually appreciate the clusterings obtained, Figure \ref{fig:osusum} presents a sample of two leaves in each of the six groups obtained by the best clustering. We can see that the two leaves in each group are aligned and present very similar shapes, which confirms that our clustering procedure is doing a good job.

\section{Conclusions}
\label{sec:concl}

We have efficiently solved the leaf clustering problem by resorting to the theory of circular densities, where scaling and rotation problems are easily solved. We further use distances between densities to produce a hierarchical clustering. As in most clustering problems, different procedures produce different groupings and the scientist has to select the clustering that better represents prior knowledge, is easier to interpret or minimises a heterogeneous measure. 

With the four distances used in the motivating dataset, we only produced two different clusterings, which by looking at the plots, both can be justified. However, for the analysis of the larger (OSU) dataset, the four distances and the two rotation mechanisms, produce different clusterings. We managed to select one by comparing with the heterogeneity measure defined in Section \ref{sec:osu}. 

The fourth distance $D_4$, which is based on the $2r$ trigonometric moments and the euclidean distance, can be considered as the benchmark which is equivalent to use normalised polar Fourier coefficients \citep{kadir&al:11}. In the analysis of the OSU Leaf dataset of Section \ref{sec:osu}, the clustering obtained with distance $D_4$ is the second best when we use the mean rotation, and the third best with the mode rotation, according to the $HM$ measure. Additionally, we observed that the resulting clusterings are very sensitive to the choice of the number of trigonometric moments $r$. 

An open aspect of our procedure is to assign a class or category to the groups obtained. We leave this to the botanists, experts in the field. 

The data analyses presented here were implemented in R \citep{r:25} and it only takes a few seconds to run. Both code and datasets are available as Supplementary Material.

\section*{Acknowledgements}

This work was supported by \textit{Asociaci\'on Mexicana de Cultura, A.C.} We are grateful to Asher Wishkerman from the Faculty of Marine Sciences at Ruppin Academic Center in Israel for calling our attention to the problem and for providing the ten plants dataset.

\bibliographystyle{natbib}

\newpage

\begin{table}
\centering
\begin{tabular}{cccccccc} \hline\hline
Distance & $G_1$ & $G_2$ & $G_3$ & $G_4$ & $G_5$ & $G_6$ & $HM$ \\ \hline
\multicolumn{8}{c}{Mean rotation} \\ \hline
$D_1$ & 36 & 37 & 45 & 10 & 49 & 23 & 4.34 \\
$D_2$ & 83 & 16 & 16 & 28 & 39 & 18 & 4.84 \\
$D_3$ & 15 & 51 & 44 & 35 & 23 & 32 & 4.57 \\
$D_4$ & 29 & 44 & 44 & 49 & 25 & 9 & 4.53 \\
\hline
\multicolumn{8}{c}{Mode rotation} \\ \hline
$D_1$ & 97 & 62 & 13 & 12 & 5 & 11 & 2.83 \\
$D_2$ & 34 & 80 & 11 & 36 & 6 & 33 & 3.37 \\
$D_3$ & 80 & 75 & 10 & 8 & 16 & 11 & 2.95 \\
$D_4$ & 53 & 75 & 19 & 16 & 31 & 6 & 3.02 \\
\hline\hline
\end{tabular}
\caption{OSU dataset. Number of elements per group $G_l$, $l=1,\ldots,6$ and $HM$ measure for each of the distances and rotations.} 
\label{tab:hm}
\end{table}

\end{document}